\documentclass[11pt]{article}

\makeatletter\renewcommand{\section}{\@startsection
{section}{1}{\z@}{-3.5ex plus -1ex minus
    -.2ex}{2.3ex plus .2ex}{\bf }}

\makeatletter\renewcommand{\subsection}{\@startsection{subsection}{2}{\z@}{-3.25ex
plus -1ex minus
   -.2ex}{1.5ex plus .2ex}{\it }}
\makeatletter\renewcommand{\subsubsection}{\@startsection{subsubsection}{3}{-2.45ex}{-3.25ex
plus -1ex minus -.2ex}{1.5ex plus .2ex}{\it }}
\renewcommand{\thesection}{\arabic{section}}
\renewcommand{\thesubsection}{\arabic{section}.\arabic{subsection}.}

\renewcommand{\theequation}{\thesection.\arabic{equation}}
\makeatletter \@addtoreset{equation}{section}

\usepackage{graphicx}
\usepackage{epsfig} 

\newcommand{\be}{\begin{equation}}
\newcommand{\ee}{\end{equation}}
\newcommand{\bea}{\begin{array}}
\newcommand{\ea}{\end{array}}
\newcommand{\beqa}{\begin{eqnarray}}
\newcommand{\eeqa}{\end{eqnarray}}
\newcommand{\nn}{\nonumber}

\renewenvironment{thebibliography}[1]
     {\baselineskip=16pt plus 2pt minus 1pt
      \section*{\large\refname
        \@mkboth{\MakeUppercase\refname}{\MakeUppercase\refname}}%
     \list{\@biblabel{\@arabic\c@enumiv}}%
           {\settowidth\labelwidth{\@biblabel{#1}}%
            \leftmargin\labelwidth
            \advance\leftmargin\labelsep
            \@openbib@code
            \usecounter{enumiv}%
            \let\p@enumiv\@empty
            \renewcommand\theenumiv{\@arabic\c@enumiv}}%
      \sloppy
      \clubpenalty4000
      \@clubpenalty \clubpenalty
      \widowpenalty4000%
      \sfcode`\.\@m}

\let\fn\footnote
\renewcommand{\footnote}[1]{\linespread{1.1}\fn{#1}\linespread{1.29}}

\usepackage{amsfonts}
\usepackage{amssymb}
\usepackage{mathrsfs}
\usepackage{amsmath,amssymb}
\usepackage{bbm}
\usepackage{bm}

\newcommand{\appendices}{\section*{Appendix}\setcounter{subsection}{0} \setcounter{equation}{0}
\renewcommand{\thesubsection}{\Alph{subsection}.}
\renewcommand{\theequation}{\thesubsection\arabic{equation}}
}

\def\tyng(#1){\hbox{\tiny$\yng(#1)$}}

\begin{document}

\begin{titlepage}
\begin{flushright}
\end{flushright}
\vskip 2.0cm

\begin{center}

\centerline{{\Large \bf Noncommutative Vortices and Flux-Tubes from Yang-Mills Theories with }} 
\vskip 1em
\centerline{{\Large \bf Spontaneously Generated Fuzzy Extra Dimensions}}

\vskip 2em

\centerline{\large \bf Se\c{c}kin~K\"{u}rk\c{c}\"{u}o\v{g}lu}

\vskip 2em

\centerline{\sl Middle East Technical University,} \centerline{\sl Department of Physics,}
\centerline{\sl In\"{o}n\"{u} Boulevard, 06531, Ankara, Turkey}

\vskip 1em

{\sl  e-mail:}  \hskip 2mm {\sl kseckin@metu.edu.tr} 
\end{center}

\vskip 2cm

\begin{quote}
\begin{center}
{\bf Abstract}
\end{center}
\vskip 2em

We consider a $U(2)$ Yang-Mills theory on ${\cal M} \times S_F^2$ where ${\cal M}$ is an arbitrary noncommutative manifold and $S_F^2$ is a fuzzy sphere spontaneously generated from a noncommutative $U(\cal {N})$ Yang-Mills theory on ${\cal M}$ coupled to a triplet of scalars in the adjoint of $U({\cal N})$. Employing the ${\rm SU(2)}$-equivariant gauge field constructed in \cite{Seckin-Derek} we perform the dimensional reduction of  the theory over the fuzzy sphere. The emergent model is a noncommutative $U(1)$ gauge theory coupled adjointly to a set of scalar fields. We study this model on the Groenewald-Moyal plane ${\cal M} = {\mathbb R}^{2}_\theta$ and find that, in certain limits, it admits noncommutative, non-BPS vortex as well as flux-tube (fluxon) solutions and discuss some of their properties. 

\vskip 5pt

\end{quote}

\vskip 1cm
\begin{quote}
\end{quote}

\end{titlepage}

\setcounter{footnote}{0}

\newpage

\section{Introduction}

Dimensional reduction of Yang-Mills theories over coset spaces of the form $G/H$ has long been an interesting theme in modern particle physics. It was first formulated in a systematic manner by Forgacs and Manton \cite{Forgacs}. The essential idea in this context can be clearly illustrated by considering a Yang-Mills theory over ${\cal M} \times G/H$, where ${\cal M}$ is a given manifold. $G$ has a natural action on its coset, and requiring the Yang-Mills gauge fields to be invariant under the $G$ action up to a gauge transformation leads to the dimensional reduction of the theory after integrating over the coset space $G/H$. This technique is usually called  ``coset space dimensional reduction'' (CSDR), and it has been widely used as a method in attempts to obtain the standard model on the Minkowski space $M^4$ starting from a Yang-Mills-Dirac theory on the higher dimensional space $M^4 \times G/H$; for a review on this topic see \cite{Zoupanos}. The widely known, prototype example of CSDR is the $SU(2)$-equivariant reduction of the Yang-Mills theory over $\mathbb{R}^4$ to an abelian Higgs model on a two-dimensional hyperbolic space $\mathbb{H}^2$, which was formulated by Witten \cite{Witten} prior to the development of the formal approach of \cite{Forgacs}, and it lead  to the construction of instanton solutions with charge greater than $1$. In this example, the coset space is a two-sphere $S^2 \equiv SU(2)/U(1)$ and $\mathbb{H}^2$ naturally appears due to the conformal equivalence of  $\mathbb{R}^4\backslash \mathbb{R}^2$ to $\mathbb{H}^2\times S^2$ together with the conformal invariance of the Yang-Mills theory in four dimensions. 

CSDR techniques have also been applied to Yang-Mills theories over ${\mathbb R}^{2d}_\theta \times S^2$ \cite{Lechtenfeld:2003cq}, where ${\mathbb R}^{2d}_\theta$ is the $2d$ dimensional Groenewald-Moyal space; a prime example of a noncommutative space. In this framework, Donaldson-Uhlenbeck-Yau (DUY) equations of a $U(2k)$ Yang-Mills theory have been reduced to a set of equations on ${\mathbb R}^{2d}_\theta$ whose solutions are given by BPS vortices on ${\mathbb R}^{2d}_\theta$ and the properties of the latter have been elaborated.

Another approach, parallel to the CSDR scheme, using the language of vector bundles and quivers is also known in the literature \cite{Prada}. In recent times, this approach has been employed in a wide variety of problems, including the formulation of quiver gauge theory of non-Abelian vortices over ${\mathbb R}^{2d}_\theta$ corresponding to instantons on ${\mathbb R}^{2d}_\theta \times S^2$, ${\mathbb R}^{2d}_\theta \times S^2 \times S^2$ \cite{Popov-Szabo, Lechtenfeld1}, to the construction of vortex solutions over Riemann surfaces which become integrable for appropriate choice of the parameters \cite{Popov} and to the construction of non-Abelian monopoles over  ${\mathbb R}^{1,1} \times S^2$ in \cite{Popov2}. In  \cite{Dolan-Szabo}, reduction of the Yang-Mills-Dirac theory on $M \times S^2$ is considered with a particular emphasis on the effects of the non-trivial monopole background on the physical particle spectrum of the reduced theory. Dimensional reduction over quantum sphere is recently studied and lead to the formulation of q-deformed quiver gauge theories and non-Abelian q-vortices \cite{Landi-Szabo}.

On another front, there have been significant advances in understanding the structure of gauge theories possessing 
fuzzy extra dimensions (for a review on fuzzy spaces see \cite{Book}). Gauge theories with fuzzy extra dimensions
using CSDR scheme were first studied in \cite{Aschieri:2003vy}. Later on this was followed by \cite{Aschieri:2006uw}, where it was shown that an $SU({\cal N})$ Yang-Mills theory on the four dimensional Minkowski space $M^4$ coupled to an appropriate set of scalar fields develops fuzzy extra dimensions in the form of fuzzy spheres $S_F^2$ via spontaneous symmetry breaking. The vacuum expectation values (VEV) for the scalar fields form the fuzzy sphere, while the fluctuations around this vacuum are interpreted as gauge fields over $S_F^2$. The resulting theory can therefore be viewed as a gauge theory over $M^4 \times S_F^2$ with a smaller gauge group; which is further corroborated by the expansion of a tower of Kaluza-Klein modes of the gauge fields over $M^4 \times S_F^2$. Inclusion of the fermions into this theory was considered in \cite{Steinacker2}, where an appropriate set of fermions in $6D$ allowed for an effective description of Dirac fermions on $M^4 \times S_F^2$, which is further affirmed by a Kaluza-Klein modes expansion over 
$S_F^2$. It was also found that a chirality constraint on the fermions leads to a description in terms of "mirror fermions"
in which each chiral fermion comes with a partner with opposite chirality and quantum numbers. Gauge theory on $M^4 \times S_F^2 \times S_F^2$ has recently been investigated in \cite{Steinacker3}. For this purpose, an $SU({\cal N})$ gauge theory on $M^4$, with the same field content as the $N=4$ SUSY Yang-Mills theory together with a potential breaking the $N=4$ supersymmetry is considered. The latter leads to the identification of the VEV's of the scalars with $S_F^2 \times S_F^2$ and the fluctuations around this vacuum as gauge fields on $S_F^2 \times S_F^2$. More recently, it was shown that twisted fuzzy spheres can be dynamically generated as extra dimensions starting from a certain orbifold projection of a N = 4 SYM theory whose consequences have been discussed in \cite{Zoupanos-1}. For a review on these results \cite{Zoupanos-Review} can be consulted. We also would like to mention that, in \cite{Lizzi} starting from a suitable matrix gauge theory, noncommutative gauge theories on ${\mathbb R}^{4}_\theta$ possessing extra dimensions have been proposed. Extra dimensions are interpreted as scalars on ${\mathbb R}^{4}_\theta$ coupled to the gauge fields, and it was shown that scalars could take vacuum expectation values leading to their identification as fuzzy spheres. Consequently, spontaneous symmetry breaking in a particular gauge theory has been investigated, and its content is compared with that of the standard model. 

In a recent article together with D. Harland \cite{Seckin-Derek}, we have addressed the question of dimensional reduction of gauge theories over fuzzy coset spaces. For this purpose, we have considered a $U(2)$ Yang-Mills theory over ${\cal M} \times S_F^2$, where ${\cal M}$ is a Riemannian manifold and $S_F^2$ is the fuzzy sphere. We performed the equivariant reduction of this model over $S_F^2$ by applying the well-known CSDR techniques and obtaining the most general $SU(2)$-equivariant gauge field over ${\cal M} \times S_F^2$. This allowed us to trace over the fuzzy sphere and thereby reduce the theory over $S_F^2$ in full. We have shown that for ${\cal M} = {\mathbb R}^2$ the emergent theory has vortex solutions depending on the parameters in the model, which correspond to instantons in the original theory. We have found that these vortices are non-BPS solutions and discussed some of their properties.    

In the present article, we continue our investigations along the lines of  \cite{Seckin-Derek} and explore a situation where the space ${\cal M}$ is also noncommutative. In this framework, we consider a $U(2)$ Yang-Mills theory ${\cal M} \times S_F^2$ where ${\cal M}$ is an arbitrary noncommutative manifold, which later on will be specified as the Groenewald-Moyal space ${\mathbb R}^{2}_\theta$. Employing the ${\rm SU(2)}$-equivariant gauge field construction of \cite{Seckin-Derek}, we perform the dimensional reduction of the theory over the fuzzy sphere. The emergent model is a noncommutative $U(1)$ gauge theory coupled adjointly to a set of scalar fields, contrary to the initial expectations that the model possesses $U(1) \times U(1)$ noncommutative gauge symmetry together with a bi-fundemental matter field due to the results obtained earlier in the equivariant reduction of the Yang-Mills theories on ${\mathbb R}^{2d}_\theta \times S^2$ in \cite{Lechtenfeld:2003cq}. In contrast, we find that the presence of extra degrees of freedom in the $SU(2)$-equivariant gauge field on $S_F^2$ leads here to a further symmetry breaking in the reduced action which turns out to be gauge invariant only if the left and the right gauge fields are identified. We study the reduced model on the Groenewald-Moyal plane ${\cal M} \equiv {\mathbb R}^{2}_\theta$ and find that, in certain limiting cases, it admits noncommutative vortex \cite{Harvey, Jatkar:2000ei, Bak} as well as flux-tube (fluxon) \cite{Polychronakos, Aganagic:2000mh} solutions which are non-BPS and devoid of a smooth commutative limit as $\theta \rightarrow 0$. In particular, we find the leading order correction in the fuzzy sphere level $\ell$ to the value of the noncommutative vortex action on ${\mathbb R}^{2}_\theta$  evaluated on the solutions (or to the energy of the static vortex when considered on ${\mathbb R}^{2}_\theta \times {\mathbb R}^1$, with ${\mathbb R}^1$ standing for time) when the fuzzy gauge constraint controlling the behavior of the radial component of the gauge field on $S_F^2$ is imposed in full and solved to leading order for the extra scalar degrees of freedom, which may be viewed as the decedents of the radial gauge field component in the reduced action. It turns out that leading corrections are of the order ${\ell^{-2}}$ and contribute to increase the vortex energy.    

Our work in the rest of the paper is organized as follows. In section 2, we give the basics of the gauge theory over ${\cal M} \times S_F^2$ and indicate how the gauge theory over ${\cal M}$ dynamically develops fuzzy sphere as extra dimensions. This is followed by a short account of the construction of the ${\rm SU(2)}$-equivariant gauge field on ${\cal M} = {\mathbb R}^2$. In section 4, we present the results of the equivariant reduction over ${\cal M} \times S_F^2$ and give the reduced action in full, which is ensued by the discussion of the structure of the reduced action and its gauge symmetry. In section 5, we present non-trivial solutions of the reduced action on ${\mathbb R}^{2}_\theta$ for two different limiting cases and discuss their properties. Extensions of our results to ${\mathbb R}^{2d}_\theta$ are also briefly given. We close with some conclusions and comments. 
  
\section{Yang-Mills Theory on ${\cal M} \times S_F^2$}
\label{sec2}

In this section, we collect the essential features of gauge theory on ${\cal M} \times S_F^2$.  We start with considering the following $U(\cal{N})$ Yang-Mills theory over a suitable noncommutative space ${\cal M}$ which we leave unspecified for the time being:
\be
S = \int_{{\cal M} } \, \mbox{Tr}_{{\cal N}} \Big (
\frac{1}{4g^2} F_{\mu \nu}^\dagger F_{\mu \nu} + 
(D_\mu \phi_a)^\dagger (D_\mu \phi_a) \Big ) + \frac{1}{\tilde{g}^2}V_1(\phi) +  a^2 V_2(\phi) \,.
\label{eq:actionfirst}
\ee
Here,  $A_\mu$ are $u({\cal N})$ valued anti-Hermitian gauge fields,  $\phi_a \,(a=1,2,3)$ are $3$ anti-Hermitian scalars
transforming in the adjoint of ${\rm SU(}{\cal N}{\rm )}$ and the covariant derivative is $D_\mu \phi_a  = \partial_\mu \phi_a + \lbrack A_\mu \,, \phi_a \rbrack$.
We take the potentials of the form 
\be 
V_1(\phi) = \, \mbox{Tr}_{{\cal N}} \big ( F_{ab}^\dagger F_{ab} \big ) \, , \quad  V_2(\phi) = \, \mbox{Tr}_{{\cal N}} \big ( (\phi_a \phi_a + {\tilde b})^2 \big ) \,.
\ee
\label{eq:potentials}
where in $V_1(\phi)$ we have defined
\be
F_{ab} := \lbrack \phi_a \,, \phi_b \rbrack - \varepsilon_{abc} \phi_c \,, 
\label{eq:curvaturefuzzy}
\ee
whose purpose will become evident shortly.

In the expressions above $a$, $\tilde{b}$, $g$ and $\tilde{g}$ are constants and $\mbox{Tr}_{{\cal N}} = {\cal N}^{-1} \mbox{Tr}$ denotes a normalized trace. We further note that $\phi_a$ transform in the vector representation of an additional global $SO(3)$ symmetry and that $V_1$ and $V_2$ are invariant under this symmetry.

As its commutative counterpart \cite{Aschieri:2006uw}, this theory spontaneously develops extra dimensions in the form of fuzzy spheres. Following \cite{Aschieri:2006uw}, let us very briefly see how this actually comes about. We observe that the potential 
$\tilde{g}^{-2}V_1 +  a^2 V_2$ is positive definite, and that solutions of
\be
F_{ab} = \lbrack \phi_a \,, \phi_b \rbrack - \varepsilon_{abc} \phi_c
= 0 \,, \quad - \phi_a \phi_a = {\tilde b}
\label{eq:minimum1}
\ee
are evidently a global minima. Most general solution to this equation is not known. However depending on the values taken by the parameter ${\tilde b}$, a large class of solutions has been found in \cite{Aschieri:2006uw}. Here we restrict ourselves to the simplest situation and refer the reader to \cite{Aschieri:2006uw} for a general discussion and its physical consequences.

Taking the value of ${\tilde b}$ as the quadratic Casimir of an irreducible representation of ${\rm SU(2)}$
labeled by $\ell$, ${\tilde b} = \ell (\ell + 1)$ with $2\ell\in\mathbb{Z}$ and assuming further that the dimension 
${\cal N}$ of the matrices $\phi_a$ is $(2 \ell +1) n$, (\ref{eq:minimum1}) is solved by the
configurations of the form 
\be
\phi_a = X_a^{(2 \ell + 1)} \otimes {\bf 1}_n \,,
\label{eq:minimumsol}
\ee
where $X_a^{(2 \ell + 1)}$ are the (anti-Hermitian) generators of ${\rm SU(2)}$ in the irreducible representation $\ell$, 
which has dimension $2\ell+1$. We observe that this vacuum configuration spontaneously breaks the ${\rm U}({\cal N})$ 
down to ${\rm U}(n)$ which is the commutant of $\phi_a$ in (\ref{eq:minimumsol}).

Fluctuations about the vacuum (\ref{eq:minimumsol}) may be written as
\be
\phi_a = X_a + A_a \,, 
\label{eq:config1} 
\ee
where $A_a \in u(2\ell+1)\otimes u(n)$ and we have used the short-hand notation $X_a^{(2 \ell + 1)} \otimes {\bf 1}_n =: X_a$.  
Then $A_a$ $(a=1,2,3)$ may be interpreted as three components of a ${\rm U}(n)$ gauge field on the fuzzy sphere $S_F^2$.
A short definition of the fuzzy sphere and some of its properties are given in Appendix A.
Thus, $\phi_a$ are the ``covariant coordinates'' on $S_F^2$ and (\ref{eq:curvaturefuzzy}) defines the associated
curvature $F_{ab}$. The latter may be expressed in terms of the gauge fields $A_a$ as:
\be
F_{ab} = \lbrack X_a \,, A_b \rbrack - \lbrack X_b \,, A_a \rbrack + \lbrack A_a \,, A_b \rbrack 
- \varepsilon_{abc} A_c \,.
\ee

Obviously, the term $V_1$ corresponds to the Yang-Mills action on $S_F^2$. However, with this term alone, gauge theory on the sphere is not recovered in the commutative limit, since the fuzzy gauge field has three components rather than two.  Rather, one obtains gauge theory with an additional scalar; the scalar is more precisely the component of the gauge field pointing in the radial direction when $S^2$ is embedded in $\mathbb{R}^3$. The purpose of the term $V_2$ in the action is to suppress this scalar.  
To see how this works, observe that
\begin{multline}
i (\ell(\ell+1))^{-1/2} \Big( (X_a+A_a)(X_a+A_a)+\ell(\ell+1) \Big) = \{ \hat{x}_a, A_a \} + i (\ell(\ell+1))^{-1/2}A_a^2 \\
\xrightarrow[ \ell\rightarrow\infty]{}  2x_a A_a \,.
\end{multline}
The term $x_aA_a$ is precisely the component of the gauge field on the sphere associated with the radial direction, so the term 
$a^2V_2$ gives a mass $a\sqrt{\ell(\ell+1)}$ to this component.

To summarize, with (\ref{eq:config1}) the action in (\ref{eq:actionfirst}) takes the form of a ${\rm U(n)}$ gauge 
theory on ${\cal M} \times S_F^2(2 \ell + 1)$ with the gauge field components $A_M({\hat y}) = (A_\mu({\hat y}) \,, A_a({\hat y}))\in  u(n)\otimes u(2\ell+1)$ and field strength tensor (${\hat y}$ are a set of coordinates for the noncommutative manifold ${\cal M}$)
\begin{eqnarray}
F_{\mu\nu} &=& \partial_\mu A_\nu - \partial_\nu A_\mu + [A_\mu,A_\nu]  \nn \\
F_{\mu a} &=& D_\mu \phi_a = \partial_\mu \phi_a + [A_\mu, \phi_a ] \\
F_{ab} &=& [\phi_a,\phi_b] - \epsilon_{abc}\phi_c \nn \,.
\end{eqnarray}

For future use, we note that
\be
\mbox{Tr}_{{\cal N}} = \frac{1}{n (2 \ell +1)} \mbox {Tr}_{\mbox{Mat} (2 \ell +1)} \otimes \mbox {Tr}_{\mbox{ Mat}(n)}
\ee
where $\mbox{Mat}(k)$ denotes the algebra of $k \times k$ matrices.

\section{The ${\rm SU(2)}$-Equivariant Gauge Field}
\label{sec3}

Let us focus on the case of a ${\rm U(2)}$ gauge theory on ${\cal M}\times S_F^2$. The
construction of the most general ${\rm SU(2)}$-equivariant gauge field on  $S_F^2$ 
was given in a recent article by the author with D. Harland \cite{Seckin-Derek}. This construction
uses essentially the representation theory of ${\rm SU(2)}$. Here we give a brief account
for completeness and refer the reader to \cite{Seckin-Derek} for further details.

We begin by selecting 
\be
\omega_a = X_a^{(2\ell+1)} \otimes {\bf 1}_2 - {\bf 1}_{2 \ell +1} \otimes \frac{i\sigma^a}{2} \,, \quad \omega_a\in u(2)\otimes u(2\ell+1) \,, \mbox{for} \,  a=1,2,3
\ee
These $\omega_a$ are the generators of the representation
$\underline{1/2} \otimes \underline{\ell}$ of ${\rm SU(2)}$, where by
$\underline{m}$ we denote the spin $m$ representation of ${\rm SU(2)}$ of
dimension $2m+1$.  The two terms which make up $\omega_a$ generate rotations and gauge transformations, therefore imposing $\omega$-equivariance amounts to requiring that rotations can be compensated by gauge transformations. 
There are certainly more possible choices for $\omega_a$; for example, $\omega_a=X_a^{(2\ell+1)} \otimes {\bf 1}_2$ was studied in \cite{Aschieri:2003vy}.

${\rm SU(2)}$-equivariance of the theory requires the fulfillment of the symmetry constraints,
\be
\lbrack \omega_a \,, A_\mu \rbrack = 0  \,,
\label{eq:A}
\ee
\be
\lbrack \omega_a, \phi_b \rbrack = \epsilon_{abc} \phi_c,
\label{eq:vector}
\ee
on the gauge field. These constraints are consistent only if
$\omega_a$ satisfies
\be
\lbrack \omega_a, \omega_b \rbrack = \varepsilon_{abc} \omega_c \,,
\ee
which is readily satisfied by our choice of $\omega_a$.

The l.h.s. of both (\ref{eq:A}) and (\ref{eq:vector}) require that 
$A_\mu$ and $\phi_a$ transform under the adjoint action of
$\omega_a$, that is, in the representation $(\underline{1/2}
\otimes \underline{\ell}) \otimes (\underline{1/2} \otimes
\underline{\ell})$ of $su(2)$. The r.h.s. of (\ref{eq:A}) and (\ref{eq:vector})
indicate that $A_\mu$ belongs to the trivial sub-representation 
and $\phi_a$ belongs to the vector sub-representation of the representation
$(\underline{1/2} \otimes \underline{\ell}) \otimes (\underline{1/2}
\otimes \underline{\ell})$.

Using the Clebsch-Gordan formula to find the sub-representations for $\ell>1/2$, we get
\begin{eqnarray}
(\underline{1/2} \otimes \underline{\ell}) &\otimes& (\underline{1/2}
\otimes \underline{\ell}) \nn \\
&=& (\underline{\ell+1/2} \oplus \underline{\ell-1/2}) \otimes
(\underline{\ell+1/2} \oplus \underline{\ell-1/2}) \\
&=& (\underline{\ell+1/2}\otimes\underline{\ell+1/2}) \oplus
2(\underline{\ell+1/2}\otimes\underline{\ell-1/2}) \oplus
(\underline{\ell-1/2}\otimes\underline{\ell-1/2}) \nn \\
&=& 2\,\underline{0} \oplus 4\,\underline{1} \oplus\dots \nn 
\end{eqnarray}

Thus, the set of solutions to (\ref{eq:A}) is two-dimensional and that of (\ref{eq:vector}) is four-dimensional.
Convenient parametrizations may be given by 
\be
A_\mu = \frac{1}{2}Q a_\mu({\hat y}) + \frac{1}{2}i b_\mu({\hat y})
\label{eq:amu}
\ee
\begin{gather}
\phi_a = X_a + A_a \,, \nn \\
A_a = \frac{1}{2}\varphi_1({\hat y})[X_a,Q] + \frac{1}{2} (\varphi_2({\hat y})-1) Q[X_a,Q] 
+ i \frac{1}{2} \varphi_3({\hat y}) \frac{1}{2} \{ \hat{X}_a, Q \} +
\frac{1}{2} \varphi_4({\hat y}) \hat{\omega}_a,,
\label{eq:eqvansatz}
\end{gather}
where  $a_\mu$, $b_\mu$ are Hermitian ${\rm U(1)}$ gauge fields, $\varphi_i$ are Hermitian scalar fields over ${\cal M}$, 
the curly brackets denote anti-commutators throughout, and we have used
\be
\hat{X}_a := \frac{1}{\ell+1/2} X_a \,, \quad {\hat \omega}_a := \frac{1}{\ell+1/2} \omega_a.
\ee
We have further introduced the anti-Hermitian matrix
\begin{equation}
Q := \frac{X_a\otimes\sigma^a - i/2}{\ell+1/2} \,, \quad Q^\dagger = - Q \,,
\quad Q^2 = - {\bf 1}_{2(2 \ell +1)} \,.
\end{equation}
Indeed, $Q$ is the fuzzy version of $q := i {\bf \sigma} \cdot {\bf x}$ and
converges to it in the $\ell \rightarrow \infty$ limit\footnote{ $\pm i Q$ appears
also in the context of monopoles and fermions over $S_F^2$ where 
in the former it is the idempotent associated with the projector describing
the rank $1$ monopole bundle over $S_F^2$, while in the latter it serves as the
chirality operator associated with the Dirac operator on $S_F^2$. For
further details on these, see for instance \cite{Book} and the references therein.}.
  
It is worthwhile to remark that, in the commutative limit ${\cal M} \rightarrow M$, $S_F^2 \rightarrow S^2$ (\ref{eq:eqvansatz}) be
comes
\be
A_a \xrightarrow[\ell \rightarrow \infty]{} i \frac{1}{2}\varphi_1(y){\cal L}_a q  + i \frac{1}{2} (\varphi_2(y)-1) q {\cal L}_a q 
+ \frac{1}{2} \varphi_3(y) x_a q + \frac{1}{2} \varphi_4(y) x_a \,.
\label{ansatz}
\ee
In this limit, the component of $A_a$ normal to $S^2$ can be eliminated by imposing the constraint 
$x_a A_a = 0$ on the gauge field.  This constraint is satisfied if and only if we take $\varphi_3 =0 \,, \varphi_4=0$, 
as is easily observed from the above expression. Thus, we recover then the well-known expression for the 
spherically symmetric gauge field over ${\cal M} \times S^2$ \cite{Witten, Forgacs}.

\section{Reduction of the Yang-Mills Action over $S_F^2$ }
\label{sec4}

Using the ${\rm SU(2)}$-equivariant gauge field in the noncommutative $U(2)$ Yang-Mills theory on ${\cal M} \otimes S_F^2$, 
we can explicitly trace it  over the fuzzy sphere to reduce it to a theory on ${\cal M}$. It is quite useful to note the following identities
\begin{gather}
\label{identity1}
\lbrace Q \,, \lbrack X_a \,, Q \rbrack \rbrace = 0 \,, \quad  \lbrace X_a \,, \lbrack X_a \,, Q \rbrack \rbrace = 0 \,,
\quad (\mbox{sum over repeated $a$ is implied}) \,, \\
\label{identity2}
\lbrack Q \,, \lbrace X_a \,, Q \rbrace \rbrack = 0 \,, \quad  \lbrack X_a \,, \lbrace X_a \,, Q \rbrace \rbrack = 0 \,,
\quad (\mbox{sum over repeated $a$ is implied}) \,.
\end{gather}
which significantly simplify the calculations, since they greatly reduce the number of traces to be computed.

The reduced action has the form
\be
S =   \int_{\cal M} {\cal L}_F + {\cal L}_G +  \frac{1}{\tilde{g}^2} V_1 +  a^2 V_2
\label{eq:reduced}
\ee
Each term in this expression is defined and evaluated below, while some details are relegated to the Appendix B. 

\subsection{The Field Strength Term}

Let us define the combinations
\be
c_\mu^\pm := \frac{1}{2} (b_\mu \pm a_\mu) \,, \quad  c_\mu^{\pm \dagger} = c_\mu^\pm \,, \quad
a_\mu = c_\mu^+ -  c_\mu^- \,, \quad \quad  b_\mu = c_\mu^+ + c_\mu^- \,.
\ee
The associated field strengths are
\be
F_{\mu \nu}^\pm = \partial_\mu c_\nu^\pm - \partial_\nu c_\mu^\pm + i \lbrack c_\mu^\pm \,, c_\nu^\pm \rbrack \,, 
\quad F_{\mu \nu }^{\pm \dagger} =  F_{\mu \nu }^\pm \,.
\ee
The corresponding contribution to the Lagrangian can then be expressed as
\begin{eqnarray}
{\cal L}_F &:=& \frac{1}{4 g^2} \mbox{Tr}_{{\cal N}} \Big ( F_{\mu \nu}^\dagger F_{\mu \nu} \Big ) \nn \\
&=& \frac{1}{4 g^2}  \frac{1}{2 \ell +1} \Big ( \ell  \left | F_{\mu\nu}^+ \right |^2  +   (\ell +1)  \left | F_{\mu\nu}^- \right |^2 \Big ) \,.
\label{eq:strength}
\end{eqnarray}

\subsection{The Gradient Term}

The covariant derivative may be written as 
\be
D_{\mu}\phi_a = \frac{1}{4} \left( ( D_\mu \varphi + D_\mu \varphi^\dagger )   -  i Q ( D_\mu
\varphi - D_\mu \varphi^\dagger ) \right )[X_a,Q] + i \beta_\mu \{ \hat{X}_a, Q \} +  \gamma_\mu \hat{\omega}_a \,.
\label{eq:gradient1}
\ee
where
\beqa
&&\varphi = \varphi_1 + i \varphi_2 \,, \quad \varphi^\dagger = \varphi_1 - i \varphi_2 \,, \nn  \\
&&D_\mu \varphi = \partial_\mu \varphi + i c_\mu^+ \varphi - i \varphi c_\mu^- \,, \quad 
D_\mu \varphi^\dagger = \partial_\mu \varphi^\dagger + i c_\mu^+ \varphi^\dagger - i \varphi^\dagger c_\mu^- \,,  \\
&&\beta_\mu = \frac{1}{4} \partial_\mu \varphi_3 + \frac{i}{8} \lbrack c_\mu^+ +c_\mu^- \,, \varphi_3 \rbrack 
+ \frac{i}{16} \left \lbrack c_\mu^+ - c_\mu^- \,, \frac{1}{\ell +\frac{1}{2}} \varphi_3 - 
\frac{(\ell + \frac{1}{2})^2- \frac{5}{2}}{(\ell + \frac{1}{2})^2 -1} \varphi_4 \right \rbrack \,, \nn \\
&&\gamma_\mu = \frac{1}{2} \partial_\mu \varphi_4 + \frac{i}{4} \lbrack c_\mu^+ +c_\mu^- \,,  \varphi_4 \rbrack -
\frac{i}{4 (\ell + \frac{1}{2})} \left \lbrack  c_\mu^+ - c_\mu^-  \,, \frac{\ell(\ell + 1)}{\ell + \frac{1}{2}} \varphi_3 
+ \frac{(\ell + \frac{1}{2})^2 - \frac{5}{8}}{(\ell + \frac{1}{2})^2 - 1} \varphi_4 \right \rbrack \,. \nn
\label{eq:1234}
\eeqa
The gradient term takes the form
\begin{multline}
{\cal L}_G := \mbox{Tr}_{{\cal N}} \Big ( (D_\mu \phi_a)^\dagger (D_\mu \phi_a) \Big ) \\
= \frac{1}{4} \frac{\ell^2+\ell}{(\ell+1/2)^2} \left( D_\mu \varphi D_\mu \varphi^\dagger + D_\mu
\varphi^\dagger D_\mu \varphi \right) 
+ 2 \frac{\ell^2+\ell}{(\ell+1/2)^2} \left ( \frac{(\ell +\frac{3}{2}) (\ell - \frac{1}{2})}{(\ell +\frac{1}{2})^2} + 1 \right ) 
\beta_\mu \beta_\mu \\
+ \frac{\ell^2 + \ell +\frac{3}{4}}{(\ell +\frac{1}{2})^2} \gamma_\mu \gamma_\mu + 2 \frac{\ell^2 + \ell}{(\ell +\frac{1}{2})^3}
\lbrace \beta_\mu \,, \gamma_\mu \rbrace \,.
\label{eq:gradient2}
\end{multline}

\subsection{The Potential Term}

Working with the dual of the curvature $F_{ab}$ we have  
\begin{multline}
\frac{1}{2} \varepsilon_{abc} F_{ab} = \frac{1}{2} \epsilon_{abc} \lbrack \phi_a,\phi_b \rbrack -\phi_c \,
= \frac{1}{2} \Big ( \lbrace P_1 \,, \varphi_1+ Q \varphi_2 \rbrace + i \lbrack S \,, Q (\varphi_1+ Q \varphi_2 ) \rbrack
\Big ) [X_c,Q] \\
+ \frac{i}{4} \left ( \varphi_1^2 + \varphi_2^2 + \frac{i}{2 (\ell +\frac{1}{2})} \lbrack \varphi_1 \,, \varphi_2 \rbrack - P_2 \right) 
\frac{ \{X_c,Q\} }{(\ell+1/2)}  + \frac{1}{4} \left ( P_3 - \frac{2i \ell(\ell +1)}{(\ell +\frac{1}{2})} \lbrack \varphi_1 \,, \varphi_2 \rbrack \right )
\frac{\omega_c}{(\ell+1/2)^2} \,,
\label{eq:dualcurvature}
\end{multline}
where $P_{1,2,3}$ and $S$ are given in the appendix B. The potential term in the action may then be expressed as
\begin{multline}
V_1 = 4 \frac{\ell^2+\ell}{(\ell+1/2)^2} (T_1^2 + T_2^2) + 4 \frac{\ell^2+\ell}{(\ell+1/2)^2} 
\left ( \frac{(\ell +\frac{3}{2}) (\ell - \frac{1}{2})}{(\ell +\frac{1}{2})^2} + 1 \right ) T_3^2 \\
+ 2 \frac{\ell^2 + \ell +\frac{3}{4}}{(\ell +\frac{1}{2})^4} T_4^2 + 4 \frac{\ell^2+\ell}{(\ell+1/2)^4} \lbrace T_3 \,, T_4 \rbrace \,,
\label{eq:potential}
\end{multline}
and the explicit expressions for $T_{1,2,3,4}$,  in terms of $P_{1,2,3}$ and $S$, are given in the appendix B.

In the large $\ell$ limit, we find  
\begin{multline}
V_1\underset{\ell \rightarrow \infty}{=} \frac{1}{4} \Bigg ( (\varphi \varphi^\dagger)^2 + (\varphi^\dagger \varphi)^2 + 
\lbrace \varphi \varphi_3 \,, \varphi^\dagger \varphi_3  \rbrace +
\lbrace \varphi \,, \varphi^\dagger \rbrace  \left ( \varphi_3^2  + 2 (\varphi_3 - 1) \right ) + 2 (\varphi_3 - 1)^2  \\
\left \lbrace \lbrack \varphi \,, \varphi^\dagger \rbrack \,, \varphi_4 \right \rbrace + 2 \varphi_4^2 \Bigg ) \,. 
\label{eq:potential1}
\end{multline}
Let us also note that in the commutative limit this collapses to 
\be
\frac{1}{2}(|\varphi|^2+\varphi_3-1)^2 + \varphi_3^2|\varphi|^2 +
\frac{1}{2}\varphi_4^2 \,,
\ee
which is the expression found in \cite{Seckin-Derek}.

\subsection{The Constraint Term}

Following the discussion in section 2, we take ${\tilde b} = \ell (\ell +1)$. We can then write
\begin{equation}
\label{constraint}
\phi_a \phi_a + \ell(\ell+1) = R_1+R_2iQ,
\end{equation}
where $R_1$ and $R_2$ are given in the appendix B.

The constraint term in the action therefore takes the form 
\begin{eqnarray}
\label{constraint term}
V_2 &=& \Big ( R_1^2+R_2^2+\frac{1}{2 (\ell + \frac{1}{2})} \lbrace R_1 \,, R_2 \rbrace \Big ).
\label{eq:constraint}
\end{eqnarray}

\subsection{Structure of the Reduced Action}

Let us now inspect the reduced action more closely and  make some important remarks and observations that  will clarify the structure of the reduced theory. For definiteness, from now on we will consider that ${\cal M }$ is the Groenewald-Moyal (GM) plane 
${\mathbb R}_\theta^2$ (see section 5.1. for definitions and our conventions on GM plane).  First, we should understand the gauge symmetry of the reduced action. In view of the results obtained in the course of the equivariant reduction of the Yang-Mills theories on ${\mathbb R}_\theta^{2d} \times S^2$ in \cite{Lechtenfeld:2003cq}, our initial expectation before performing the dimensional reduction has been to encounter a reduced theory with a $U(1) \otimes U(1)$ gauge symmetry where $\varphi \,, \varphi^\dagger$ are in the bi-fundamental representation and $\varphi_3 \,, \varphi_4$ are neutral scalars and therefore in the adjoint of both the left and the right $U(1)$ factor; in fact the latter is the only option for  $\varphi_3 \,, \varphi_4$, since they can not be carrying any charges (except the same charges $(1,-1)$ as $\varphi$ under the left and the right gauge groups, respectively, which  they certainly do not carry, as is clear from the form of $\beta_\mu$ and $\gamma_\mu$) under  the $U(1) \otimes U(1)$ gauge group as the noncommutativity will prevent that from happening \cite{Chaichian:2001mu}, and from the form of $\beta_\mu$ and $\gamma_\mu$ it is also clear that they can not be transforming under the trivial representation of $U(1) \otimes U(1)$. Therefore, in contrast to the results of \cite{Lechtenfeld:2003cq}, where of course $\varphi_3 \,, \varphi_4$ were absent, we find that the presence of extra degrees of freedom, namely $\varphi_3 \,, \varphi_4$, in the $SU(2)$-equivariant gauge field on $S_F^2$ leads here to a further symmetry breaking in the reduced action. Inspecting the expressions (4.8) making up the ingredients of the gradient term in (\ref{eq:gradient2}), we see that the gauge symmetry is broken down to a diagonal noncommutative $U(1)$ gauge group. We observe that $\beta_\mu$ and $\gamma_\mu$ transform covariantly and the reduced action is invariant only under the diagonal noncommutative $U(1)$ gauge group, that is only if the left and the right gauge fields are identified:  $c_\mu^+ = c_\mu^- = : c_\mu$ (with this definition $c_\mu = \frac{1}{2} b_\mu$). We find that the reduced action then takes the form 
\begin{multline}
S =   \int_{\cal M} 
\frac{1}{4 g^2}  \left | F_{\mu\nu} \right |^2 + \frac{1}{2} \frac{\ell^2+\ell}{(\ell+1/2)^2}  D_\mu \varphi D_\mu \varphi^\dagger 
+ \frac{1}{8} \frac{\ell^2+\ell}{(\ell+1/2)^2} \left ( \frac{(\ell +\frac{3}{2}) (\ell - \frac{1}{2})}{(\ell +\frac{1}{2})^2} + 1 \right ) 
(D_\mu \varphi_3)^2 \\ 
+ \frac{\ell^2 + \ell +\frac{3}{4}}{4 (\ell +\frac{1}{2})^2} (D_\mu \varphi_4)^2 +  \frac{\ell^2 + \ell}{4 (\ell +\frac{1}{2})^3}
\lbrace D_\mu \varphi_3 \,, D_\mu \varphi_4  \rbrace +  \frac{1}{\tilde{g}^2} V_1 +  a^2 V_2 \,.
\label{eq:fullreduced}
\end{multline}
where now we have
\be
D_\mu \cdot = \partial_\mu \cdot + \lbrack  c_\mu \,, \cdot \rbrack \,, \quad F_{\mu \nu} = 
\partial_\mu c_\nu - \partial_\nu c_\mu + i \lbrack c_\mu \,, c_\nu \rbrack \,.
\ee

We note that the theory governed by the action (\ref{eq:fullreduced}) does not have a commutative limit, since then all the commutators vanish and the remaining terms no longer form a gauge theory. This is a well-known behavior of gauge theories with adjoint scalar fields \cite{Jatkar:2000ei, Nekrasov-Review}, and it is also encountered in the present model. However, it is also useful to remark that, taking the commutative limit in the action (\ref{eq:reduced}) first using the expressions
(\ref{eq:strength}, \ref{eq:gradient2}, \ref{eq:potential}) and (\ref{eq:constraint}) leads to the results found in \cite{Seckin-Derek}. 

\section{Solutions of the Reduced Theory on ${\mathbb R}_\theta^2$}
\label{sec6}

We now wish to study the classical solutions of the system governed by the action given in (\ref{eq:fullreduced}) on the Groenewald-Moyal
plane ${\mathbb R}_\theta^2$. As emphasized in \cite{Seckin-Derek} there is no canonical choice for the coefficient $a^2$ of the fuzzy 
constraint term; we will consider the two extreme cases of  $a^2=\infty$ and $a^2=0$ corresponding, respectively, to imposing the constraint $\phi_a\phi_a+\ell(\ell+1)=0$ in full (i.e. ``by hand'') and imposing no constraint at all. In both cases, we consider the large $\ell$ limit; in the 
$a=\infty$ theory, we include only terms appearing at $O(\ell^{-2})$, whereas for the case $a=0$, we assume $\ell=\infty$.

\subsection{Definitions and Conventions for the Groenewald-Moyal Plane ${\mathbb R}_\theta^2$}

Using the operator formalism, ${\mathbb R}_\theta^2$ is defined by two operators ${\hat y}_1 \,, {\hat y}_2$ acting on the standard Harmonic oscillator Fock space ${\cal H}$. They fulfill the Heisenberg algebra commutation relation
\be
\lbrack \hat {y}_1 \,, \hat{y}_2 \rbrack = i \theta \,,
\ee
where $\theta$ is the noncommutativity parameter.

It is often useful to switch to the complex basis which we take as
\be
z = \frac{1}{\sqrt{2}} (y_1 + i y_2)  \,, \quad \bar{z} = \frac{1}{\sqrt{2}} (y_1 - i y_2) \,,
\ee
fulfilling
\be
\lbrack z \,, \bar{z} \rbrack = \theta \,.
\ee
The derivatives on ${\mathbb R}_\theta^2$ maybe expressed as 
\be
\partial_\mu \cdot = - \frac{i}{\theta} \varepsilon_{\mu \nu} \lbrack {\hat y}_\nu \,, \cdot \rbrack \,, \quad 
\partial_z \cdot = - \frac{1}{\theta} \lbrack \bar{z} \,, \cdot \rbrack \,, \quad
\partial_{\bar z} \cdot = \frac{1}{\theta} \lbrack z \,, \cdot \rbrack \,.   
\ee
The integration over ${\mathbb R}^2$ becomes a trace over the Fock space  ${\cal H}$ on ${\mathbb R}_\theta^2$:
\be
\int_{{\mathbb R}^2} d^2 y  \longrightarrow 2 \pi \theta \, \mbox{Tr}_{{\cal H}} \,.
\ee
For further details on noncommutative spaces, see for instance \cite{Nekrasov-Review}. 

\subsection{Case 1: The constraint fully imposed}

The fuzzy constraint $\phi_a\phi_a+\ell(\ell+1)=0$ is equivalent to the two algebraic equations $R_1=0$, $R_2=0$, where the expression for $R_1$ and $R_2$ are given in the Appendix B. These equations can be solved order by order in powers of the parameter $\frac{1}{\ell}$ to obtain $\varphi_3$ and $\varphi_4$ in terms of $\varphi_1$ and $\varphi_2$.  Substituting back 
into the action yields an action involving only the scalar $\varphi=\varphi_1+i\varphi_2$.

When $\ell=\infty$, the solution to the constraint is simply $\varphi_3=0$, $\varphi_4=0$, and substituting these into the action
(\ref{eq:reduced}) yields the model found in \cite{Lechtenfeld:2003cq}, where $\varphi$ is a bi-fundemental scalar field, and there are distinct left and right gauge fields, which are not required to be identified. When finite $\ell$ effects are taken into account, however, we should consider the action (\ref{eq:fullreduced}); then gauge invariance of the actions both before and after solving the constraint, and gauge covariance of the solutions of the constraint are maintained only if the left and the right gauge fields are identified. 

For large but finite $\ell$, one can solve the constraint approximately by expanding it to leading order in powers of $\ell^{-1}$ around the $\ell=\infty$.  
Performing this to order  $O\left(\ell^{-3}\right)$, we find
\beqa
\varphi_3 &=& - i\frac{4}{\ell} \lbrack \varphi_1 \,, \varphi_2 \rbrack  + \frac{1}{2\ell^2}(\varphi_1^2 + \varphi_2^2 - 1) + O \left(\ell^{-3}\right)\,, \\
\varphi_4 &=& - \frac{1}{2\ell}(\varphi_1^2 + \varphi_2^2 - 1) + i\frac{3}{\ell^2} \lbrack \varphi_1 \,, \varphi_2 \rbrack + O \left(\ell^{-3}\right)\,.
\label{eq:consolutions}
\eeqa
Note that these indeed preserve the gauge symmetry since both sides transform covariantly under the action of the gauge group.

Using these in (\ref{eq:fullreduced}), we find 
\begin{multline} 
S =   2 \pi \theta \mbox{Tr}_{{\cal H}} \Bigg \lbrack 
\frac{1}{4 g^2}  \left | F_{\mu\nu} \right |^2 + \frac{1}{2} \left (1 - \frac{1}{4 \ell^2} \right ) D_\mu \varphi D_\mu \varphi^\dagger 
+\frac{1}{\ell^2} \left ( D_\mu \lbrack \varphi \,, \varphi^\dagger \rbrack \right )^2 + 
\frac{1}{32 \ell^2} \left ( D_\mu \lbrace \varphi \,, \varphi^\dagger \rbrace \right ) ^2 \\
+ \frac{1}{{\tilde g}^2} \Bigg ( \left ( \frac{1}{2} + \frac{1}{4 \ell^2} \right ) \left ( \frac{1}{2} \lbrace \varphi \,, \varphi^\dagger \rbrace  - 1 \right )^2 +
\frac{1}{8} \left (1 - \frac{1}{\ell} - \frac{3}{4 \ell^2} \right )  \lbrack \varphi \,, \varphi^\dagger \rbrack^2 \Bigg ) + O \left(\ell^{-3}\right)  \Bigg \rbrack \,.
\label{eq:reduced-expanded-simplified}
\end{multline}
To obtain this result we have also used the cyclicity property of the trace $\mbox{Tr}_{{\cal H}}$, under which the terms 
proportional to $\lbrack \varphi \,, \varphi^\dagger \rbrack$ and  $\left \lbrace \lbrace \varphi \,, \varphi^\dagger \rbrace \,,
\lbrack \varphi \,, \varphi^\dagger \rbrack \right \rbrace $ vanish.
The expression (\ref{eq:reduced-expanded-simplified})  is clearly invariant under the noncommutative $U(1)$ gauge symmetry, as it should be.

It is possible to employ the solution generating techniques introduced in \cite{Harvey} to find
noncommutative vortex type solutions of (\ref{eq:reduced-expanded-simplified}). To this end we proceed as follows. 
Let us first define the covariant coordinates
\be
X = - \frac{1}{\theta} {\bar z} + i c_z \,, \quad X^\dagger = - \frac{1}{\theta} z - i c_{{\bar z}} \,,
\label{eq:covariantgaugef}
\ee
where we have used the complex combinations $c_z =\frac{1}{\sqrt{2}} (c_1 - i c_2)$, $c_{\bar z} = \frac{1}{\sqrt{2}} (c_1 + i c_2)$.
The covariant derivatives and the field strength may be expressed as
\be
D_z \varphi = \lbrack X \,, \varphi \rbrack \,, \quad D_{\bar z} \varphi = - \lbrack X^\dagger \,, \varphi \rbrack \,.
\ee
\beqa
F_{z {\bar z}} &=& \partial_z c_{\bar z} - \partial_{\bar z} c_z + i \lbrack c_z  \,, c_{\bar z} \rbrack \,, \nn \\
&=& i \lbrack X \,, X^\dagger \rbrack + \frac{i}{\theta} \,,  
\eeqa
All the basic constituents of the action  (\ref{eq:reduced-expanded-simplified}) transform covarianty under the gauge symmetry
\be
X \longrightarrow U X U^\dagger \,, \quad \varphi \longrightarrow U \varphi U^\dagger \,, \quad
D_z \varphi \longrightarrow U D_z \varphi U^\dagger \,, \quad  F_{z {\bar z}}  \longrightarrow
U F_{z {\bar z}}  U^\dagger \,.
\ee
It follows that the equations of motion will transform covariantly, that is,
\be
\frac{\delta S}{\delta X} \longrightarrow U \frac{\delta S}{\delta X} U^\dagger \,, \quad \frac{\delta S}{\delta \varphi} \longrightarrow U \frac{\delta S}{\delta \varphi} U^\dagger \,,
\label{eq:solutiongenerator}
\ee
under a partial isometry $U$ satisfying
\be
U^\dagger U = 1 \,, \quad U U^\dagger = P \,,
\label{eq:partialiso}
\ee
where $P$ is a projection operator \cite{Harvey}. Thus, the partial isometries (\ref{eq:partialiso}) generate solutions from a known solution.

A trivial solution to the equations of motion of (\ref{eq:reduced-expanded-simplified}) may easily found  to be
$X = - \frac{1}{\theta} {\bar z} \,, \varphi = 1$. Taking $U = S^m$, where $S$ is the usual shift operator $S = \sum_{k=0}^\infty | k +1 \rangle \langle k |$, we can write a set of non-trivial solutions for the theory governed by (\ref{eq:reduced-expanded-simplified})  as
\beqa
\varphi  &=& S^m S^{\dagger m} = 1 - P_m \,, \nn \\
X &=& - \frac{1}{\theta} S^m {\bar z} S^{\dagger m} \,.
\eeqa
where 
\be
P_n = \sum_{k=0}^{n-1} | k \rangle \langle k | \,,
\ee
is the projection operator of rank $m$. The corresponding field strength is $F_{12} = - i F_{z \bar{z}} = \frac{1}{\theta} P_m$.
We can view these solutions as noncommutative vortices carrying $m$ units of flux: 
\be
2 \pi \theta \, \mbox{Tr} F_{12} = 2 \pi m \,.
\ee

It is useful to evaluate the value of the action (\ref{eq:reduced-expanded-simplified}) on these solutions; we find
\be
S = \pi \theta m \left ( \frac{1}{ g^2 \theta^2} + \frac{1}{{\tilde g}^2} \left ( 1 + \frac{1}{2 \ell^2} \right ) \right ) + O \left(\ell^{-3}\right) \,.
\label{eq:actionvalue}
\ee
This corresponds to the energy of the static vortices in $2+1$ dimensions, ${\mathbb R}^2_\theta \times {\mathbb R}^1$
with ${\mathbb R}^1$ standing for time. We observe that to leading order in $\ell^{-1}$ there is a ${\ell^{-2}}$ contribution 
adding to the energy, which is a residue of the fact that the present model has descended from a model with a fuzzy sphere of order $\ell$, $S_F^2(\ell)$ as extra dimensions.

Two limiting cases may also be easily recorded from (\ref{eq:actionvalue}). For ${\tilde g} \rightarrow \infty$, our solutions collapse to the flux- tube (fluxon) solutions discussed in \cite{Polychronakos, Aganagic:2000mh}; whereas, for  $\theta \rightarrow \infty$, the action gets a contribution only from the potential term, and our vortex solution collapses to a noncommutative soliton solution of the type first discussed in \cite{Gopakumar:2000zd}.

\subsection{Case 2: No constraint}

With $a=0$ and $\ell=\infty$, the action reduces to
\begin{multline}
S =   2 \pi \theta \mbox{Tr}_{{\cal H}} \Bigg (
\frac{1}{4 g^2}  \left | F_{\mu\nu} \right |^2 + \frac{1}{2} D_\mu \varphi D_\mu \varphi^\dagger 
+\frac{1}{4}  (D_\mu \varphi_3)^2 + \frac{1}{4} (D_\mu \varphi_4)^2 + \frac{1}{\tilde{g}^2} V_1  \Bigg ) \,.
\label{eq:fullreducedinfiniteell}
\end{multline}
where $V_1$ is as given in (\ref{eq:potential1}). We see that there are linear terms in the potential (\ref{eq:potential1}) in $\varphi_3$, which will prevent us from applying the solution generating technique used in the previous subsection since these lead to terms in the equations of motion proportional to identity, and, therefore, they do not transform adjointly under the solution generating transformations \cite{Harvey}. However, in the present model this situation can be remedied by defining a new field $\varphi_3^\prime = \varphi_3 -1$. In this manner, all the terms in the potential are quadratic or higher order or a constant. We have
\begin{multline}
V_1^\prime \underset{\ell \rightarrow \infty}{=} \frac{1}{4} \Bigg ( (\varphi \varphi^\dagger)^2 + (\varphi^\dagger \varphi)^2 + 
\lbrace \varphi (\varphi_3^\prime + 1) \,, \varphi^\dagger (\varphi_3^\prime + 1) \rbrace +
\lbrace \varphi \,, \varphi^\dagger \rbrace  \left ( (\varphi_3^\prime + 1) ^2  + 2 \varphi_3^\prime \right ) + 2 \varphi_3^{\prime 2}  \\
\left \lbrace \lbrack \varphi \,, \varphi^\dagger \rbrack \,, \varphi_4 \right \rbrace + 2 \varphi_4^2 \Bigg ) \,. 
\end{multline}
while the gradient term involving $\varphi_3$ is unaffected by this substitution. The equations of motion are
\beqa
&&(D_z D_{\bar z} + D_{\bar z} D_z ) \phi - \frac{\partial V_1^\prime}{\partial \phi} = 0 
\,, \quad \mbox{for} \, \phi: \varphi \,, \varphi_3^\prime, \varphi_4  \,, \nn \\
&& \frac{1}{g^2}D_z F_{z \bar{z}} + i (\varphi D_{\bar z} \varphi^\dagger - D_{\bar z} \varphi \varphi^\dagger )
+i \lbrack \varphi_3^\prime \,, D_{\bar z} \varphi_3^\prime \rbrack + i \lbrack \varphi_4 \,, D_{\bar z} \varphi_4 \rbrack = 0 \,.
\label{eq:eom}
\eeqa 
We observe that a trivial solution to these equations is given by $\varphi=1 \,, \varphi_3^\prime = -1 \,, \varphi_4 = 0 \,, X = - \frac{1}{\theta} {\bar z}$. Applying the solution generating technique with $U= S^m$, we find
\be
\varphi = S^m S^{\dagger m} = 1 - P_m \,, \quad  \varphi_3^\prime = - S^m S^{\dagger m}  = P_m - 1\,, \quad \varphi_4 = 0 \,,
\quad X = - \frac{1}{\theta} S^m {\bar z} S^{\dagger m} \,,
\label{eq:fluxtube}
\ee
where $S$ and $X$ are defined as in the previous subsection.

Evaluating the value of the action (\ref{eq:fullreducedinfiniteell}) on these solutions we find
\be
S = \frac{\pi m}{ g^2 \theta} \,,
\label{eq:actionvalueinfinitell}
\ee
and the flux carried by these solutions is again
\be
2 \pi \theta \, \mbox{Tr} F_{12} = 2 \pi m \,.
\ee
As it turns out, there is in fact no contribution to (\ref{eq:actionvalueinfinitell}) from the potential term. Thus, we can interpret
(\ref{eq:fluxtube}) as flux-tube solutions carrying $m$ units of flux \cite{Polychronakos}. It is easy to see that (\ref{eq:fluxtube}) satisfies the equations of motion (\ref{eq:eom}), by noting that 
\be
D_z \phi = \lbrack X \,, \phi \rbrack = 0 \,, \quad D_{\bar z} \phi = - \lbrack X^\dagger \,, \phi \rbrack = 0 \,,
\ee
where $\phi$ are the solutions for $\varphi, \varphi_3^\prime, \varphi_4$ given in (\ref{eq:fluxtube}).

We also wish to remark that the field redefinition for $\varphi_3$ used above works only in the infinite $\ell$ limit. In fact, there does not 
appear to be a field redefinition at finite $\ell$ or at leading order around $\ell = \infty$ which will allow the use of solution generating 
transformations to construct non-trivial solutions.

\subsection{Generalization to ${\mathbb R}_\theta^{2d}$}

Results of the previous sections can be generalized to ${\mathbb R}_\theta^{2d}$ in a rather straightforward manner. Defining relations for 
${\mathbb R}_\theta^{2d}$ are
\be
\lbrack \hat {y}_\mu \,, \hat{y}_\nu \rbrack = i \theta^{\mu \nu} \,,
\ee
where it is assumed that  $\theta^{\mu \nu}$ is brought to a block-diagonal form with
\be
\theta^{2a-1 \, 2a} = - \theta^ {2a \, 2a-1} = : \theta^a \,, \quad (a = 1,\cdots , d) \,.
\ee
In complex coordinates
\be
z^a = \frac{1}{\sqrt{2}} (y^{2a-1} + i y^{2a})  \,, \quad z^{\bar a} = \frac{1}{\sqrt{2}} (y^{2a-1} - i y^{2a}) \,,
\ee
these relations become
\be
\lbrack z^a \,, z^{\bar b} \rbrack = \delta ^{a \bar b} \theta^a  = \theta^{a \bar b} = - \theta^{\bar b a} \,,
\ee
with $\theta_{b \bar c} \theta^{\bar c a} = \delta^a_b$, $\theta_{a \bar b} = - \theta_{\bar b a} = - \frac{1}{\theta^a} \delta_{a \bar b}$. We further have
\be
\int_{{\mathbb R}^{2d}} d^{2d} y  \longrightarrow \left ( \prod_{a=1}^d 2 \pi \theta^a \right ) \, \mbox{Tr}_{{\cal H}} \,.
\ee

In order to write down the generalizations of our previous results, we can consider a $U(2k)$ Yang-Mills gauge theory on ${\mathbb R}_\theta^{2d} \times S_F^2$, instead of the $U(2)$ theory that we have used in section 3. In this case, the gauge fields $A_a$ are elements of $u(2k) \otimes u(2 \ell+1)$ and $SU(2)$ equivariance therefore leads to the gauge fields $a_\mu$, $b_\mu$ taking values in $u(k)$ and to the scalar fields $\varphi_1, \varphi_2, \varphi_3, \varphi_4$ which are $k \times k$ Hermitian matrices. Dimensional reduction over the fuzzy sphere proceeds in the same manner as before.

To obtain the non-trivial solutions of the reduced theory, we again make use of the covariant coordinates
\be
X_a = i c_a + \theta_{a \bar b} z^{\bar b} \,.
\ee
For the case considered in 5.2., where the constraint term is fully imposed, we find that the non-trivial solutions generalize to
\be
X_a = \theta_{a \bar b} T_m z^{\bar b} T_m^\dagger \,, \quad \varphi = \varphi^\dagger = T_m T_m^\dagger = 1 - P_m \,,
\ee
where $T_m$, $T_m^\dagger$ are $k \times k$ matrices acting on ${\mathbb C}^k \otimes {\cal H}$ and satisfying
$T_m T_m^\dagger = 1- P_m$, $ T_m^\dagger T_m = 1$. $P_m$ is a rank $m$ projector on ${\mathbb C}^k \otimes {\cal H}$.
Explicit constructions of the operators $T_m$ are given in terms of noncommutative ABS construction, which is well-known in the literature
\cite{Harvey, Lechtenfeld:2003cq}. 

Assuming for simplicity that $\theta^1 = \cdots \theta^d = \theta$ we find that the generalized static noncommutative vortices 
have the energy
\be
(2 \pi \theta )^d m  \left ( \frac{d}{ 2 g^2 \theta^2} + \frac{1}{{\tilde g}^2} \left ( 1 + \frac{1}{2 \ell^2} \right ) \right ) + O \left(\ell^{-3}\right) \,.
\ee

As for the case of section 5.3. where the constraint term is neglected by setting $a=0$, we have the non-trivial solutions
\be
\varphi= 1 - P_m \,, \quad \varphi_3^\prime = P_m - 1 \,, \quad \varphi_4 = 0 \,, \quad X_a = \theta_{a \bar b} T_m z^{\bar b} T_m^\dagger \,.
\ee 
Evaluating the action on these solutions we get
\be
S = (2 \pi \theta)^d \frac{m d}{2 g^2 \theta^2} \,.
\ee
Thus these solutions are the generalized fluxons living on ${\mathbb R}_\theta^{2d}$. 

\section{Conclusions}

In this article, we have studied the equivariant dimensional reduction of a $U(2)$ Yang-Mills theory on ${\cal M} \times S_F^2$ where 
$M$ is considered as a noncommutative manifold. We have employed ${\rm SU(2)}$-equivariant gauge field constructed in \cite{Seckin-Derek} to perform the dimensional reduction of  the theory over the fuzzy sphere in full. Our results showed that the reduced model is a
noncommutative $U(1)$ gauge theory coupled adjointly to a set of scalar fields. We have examined the reduced model
on ${\mathbb R}^{2}_\theta$ and found that, in certain limits, it admits noncommutative vortex as well as flux-tube solutions which are non-BPS and devoid of a smooth commutative limit. In particular, we have computed the leading order 
correction in the fuzzy sphere level $\ell$ to the noncommutative static vortex energy when the fuzzy gauge constraint is fully imposed.
Generalizations of our results to $U(2k)$ gauge theories over ${\mathbb R}^{2d}_\theta$ are also briefly given.
\vskip 1em

{\bf Acknowledgements}

\vskip 1em

This work is supported by the Middle East Technical University under Project No. BAP- 08-11-2010-R-108.

\appendices

\subsection{The Fuzzy Sphere}

The fuzzy sphere at level $\ell$ is defined to be the algebra of $(2\ell +1) \times (2 \ell +1)$ matrices 
$\mbox{Mat}(2 \ell +1)$.  The three Hermitian ``coordinate functions''
\be
{\hat x}_a := \frac{i}{\sqrt{\ell (\ell +1)}} X_a^{(2\ell+1)}
\ee
satisfy
\be
\lbrack {\hat x}_a \,, {\hat x}_b \rbrack = \frac{i}{\sqrt{\ell (\ell +1)}} \varepsilon_{abc} {\hat x}_c \,, 
\quad {\hat x}_a {\hat x}_a = 1 \,,
\ee
and generate the full matrix algebra $\mbox{Mat}(2 \ell +1)$.  There are three natural derivations of functions, defined by the adjoint action of $su(2)$ on $S_F^2$:
\be
f \rightarrow ad X_a^{(2 \ell + 1)} f := \lbrack X_a^{(2 \ell + 1)} \,, f \rbrack \,, \quad f \in \mbox{Mat}(2 \ell +1) \,.  
\ee
In the limit $\ell\rightarrow\infty$, the functions $\hat{x}_a$ are identified with the standard coordinates 
$x_a$ on $\mathbb{R}^3$, restricted to the unit sphere, and the infinite-dimensional algebra ${\cal C}^\infty(S^2)$ 
of functions on the sphere is recovered. Also in this limit, the derivations $[X_a^{(2 \ell + 1)},\cdot]$ become the vector 
fields $-i{\cal L}_a = \varepsilon_{abc} x_a \partial_b$, induced by the usual action of $SO(3)$.

\subsection{Explicit Formulae}

In this appendix, we list the explicit expressions for $P_1\,, P_2\,, P_3$, $S$, $T_1\,, T_2\,, T_3, T_4$ and $R_1\,, R_2$
which were introduced for brevity of notation in section 5. 

We have

\beqa
P_1 &=& \frac{1}{2} \frac{\ell^2+\ell-1/4}{(\ell+1/2)^2}\varphi_3 + \frac{1}{2} \frac{1}{\ell+1/2}\varphi_4 \,, \\
P_2 &=& \frac{1}{2} \left \lbrace 1-\varphi_3 \,,  1 + \frac{\varphi_4}{\ell+1/2} - \frac{\varphi_3}{2(\ell+1/2)^2} \right \rbrace \,,\\
P_3 &=& \frac{\ell^2+\ell}{(\ell+1/2)^2} \left( \varphi_3^2-2\varphi_3
\right) + \varphi_4^2 + 2\frac{\ell^2+\ell-1/4}{\ell+1/2} \varphi_4 \,.
\eeqa

\be
S = \frac{1}{4 (\ell + \frac{1}{2})} \varphi_3 + \frac{1}{2} \varphi_4 \,.
\ee

In terms of $P_{1,2,3}$ and $S$ we have 
\beqa
T_1 &=& \frac{1}{2} \left ( \lbrace P_1 \,, \varphi_1 \rbrace - i \lbrack S \,, \varphi_2 \rbrack \right ) \,, \\
T_2 &=& \frac{1}{2} \left ( \lbrace P_1 \,, \varphi_2 \rbrace + i \lbrack S \,, \varphi_1 \rbrack \right ) \,, \\
T_3 &=&  \frac{1}{4} \left ( \varphi_1^2 + \varphi_2^2 + \frac{i}{2 (\ell + \frac{1}{2})} \lbrack \varphi_1 \,, \varphi_2 \rbrack 
- P_2 \right ) \,, \\
T_4 &=& \frac{1}{4} \left ( P_3 - \frac{2 i \ell (\ell + 1)}{(\ell + \frac{1}{2})} \lbrack \varphi_1 \,, \varphi_2 \rbrack \right ) \,.
\eeqa

For $R_1$ and $R_2$, we have
\begin{multline}
R_1 = -\frac{1}{2}(\varphi_1^2+\varphi_2^2-1) + \frac{3 i}{\ell +\frac{1}{2}}  \lbrack \varphi_1 \,, \varphi_2 \rbrack
- \frac{1}{4(\ell + \frac{1}{2})^2}\varphi_3 - \left((\ell +
\frac{1}{2})-\frac{1}{2(\ell + \frac{1}{2})}\right) \varphi_4 \\ 
- \left( \frac{1}{4} - \frac{3}{16(\ell + \frac{1}{2})^2} \right) \varphi_3^2
- \frac{1}{8 (\ell + \frac{1}{2})} \lbrace \varphi_3 \,, \varphi_4 \rbrace - \frac{1}{4}\varphi_4^2  \,,
\end{multline}
\begin{multline}
R_2 = \frac{1}{4(\ell + \frac{1}{2})}(\varphi_1^2+\varphi_2^2-1) - i \frac{4 (\ell + \frac{1}{2})^2 - \frac{1}{2}}{ (\ell + \frac{1}{2})^2} 
\lbrack \varphi_1 \,, \varphi_2 \rbrack -\left( (\ell + \frac{1}{2}) - \frac{3}{4(\ell + \frac{1}{2})}\right)
\varphi_3 \\
- \frac{1}{2}\varphi_4 - \frac{1}{16(\ell + \frac{1}{2})^3}\varphi_3^2 - \left( \frac{1}{4} - \frac{1}{8(\ell + \frac{1}{2})^2} \right) \lbrace \varphi_3 \,, \varphi_4 \rbrace - \frac{1}{4(\ell + \frac{1}{2})}\varphi_4^2 \,.
\end{multline}

\end{document}